\documentclass[a4paper,11pt]{article}
\usepackage{subcaption}
\usepackage{comment}
\usepackage{pos}
\usepackage{xcolor}
\usepackage{xargs}
\usepackage{bbm}
\usepackage[colorinlistoftodos,prependcaption,textsize=footnotesize]{todonotes}
\newcommandx{\DM}[2][1=]{\todo[linecolor=orange,backgroundcolor=orange!25,bordercolor=orange,#1]{DM: #2}}
\newcommandx{\KB}[2][1=]{\todo[linecolor=purple,backgroundcolor=purple!25,bordercolor=purple,#1]{KB: #2}}
\newcommandx{\PH}[2][1=]{\todo[linecolor=yellow,backgroundcolor=yellow!25,bordercolor=yellow,#1]{PH: #2}}
\newcommandx{\TODO}[2][1=]{\todo[linecolor=red,backgroundcolor=red!25,bordercolor=red,#1]{TODO: #2}}

\usepackage{layouts}
\usepackage{wrapfig}
\usepackage{floatrow}

\newcommand\flowsfromb{\mathrel{\reflectbox{$\leadsto$}}}

\title{Advancing real-time Yang-Mills: towards real-time observables from first principles}
\ShortTitle{Towards real-time observables from first principles}

\author[a]{Kirill Boguslavski}
\author*[a]{Paul Hotzy}
\author[a]{David I.\ M\"uller}

\affiliation[a]{Institute of Theoretical Physics, TU Wien, Wiedner Hauptstraße 8-10, 1040 Vienna, Austria}

\emailAdd{kirill.boguslavski@tuwien.ac.at}
\emailAdd{paul.hotzy@tuwien.ac.at}
\emailAdd{dmueller@hep.itp.tuwien.ac.at}

\abstract{
The complex Langevin (CL) method shows great promise in enabling the calculation of observables for theories with complex actions. Nevertheless, real-time quantum field theories have remained largely unsolved due to the particular severity of the sign problem. In this contribution, we discuss our recent progress in applying CL to a thermal SU(2) Yang-Mills theory on a 3+1 dimensional lattice. We present our anisotropic kernel that stabilizes the CL approach for real times longer than the inverse temperature -- a first for Yang-Mills theory. We provide explicit evidence of reproducing symmetries and relations among different types of propagators when the complex time path approaches the Schwinger-Keldysh contour. This method paves the way for calculating transport coefficients and other real-time observables from first principles.
}

\FullConference{The 40th International Symposium on Lattice Field Theory (Lattice 2023)\\
July 31st - August 4th, 2023\\
Fermi National Accelerator Laboratory\\}

\begin{document}
\maketitle

\section{Introduction}

The quark-gluon plasma characterizes the state of matter of strongly interacting quarks and gluons. It is formed in the earliest instants in ultrarelativistic heavy-ion collisions in collider experiments such as RHIC and the LHC and has likely existed in the early (hot and dense) stages of our universe. 

\begin{figure}[b]
    \floatbox[{\capbeside\thisfloatsetup{capbesideposition={right,center},capbesidewidth=0.4\textwidth}}]{figure}[\FBwidth]
    {\caption{Visualization of the Schwinger-Keldysh (SK) contour in the continuum (blue) and discretized and tilted on the lattice (green). Dotted lines represent the contour with a tilt angle $\alpha$ regulating the path integral; as $\alpha$ approaches 0, it converges to the SK contour. The green line indicates the discretized contour with $\alpha=0$, where the finite angle serves for visualization purposes only. All depicted curves connect $t=0$ and $t=-i\beta$, where $\beta$ denotes the inverse temperature of the systems.}\label{fig:SK}}
    {\includegraphics[width=0.5\textwidth]{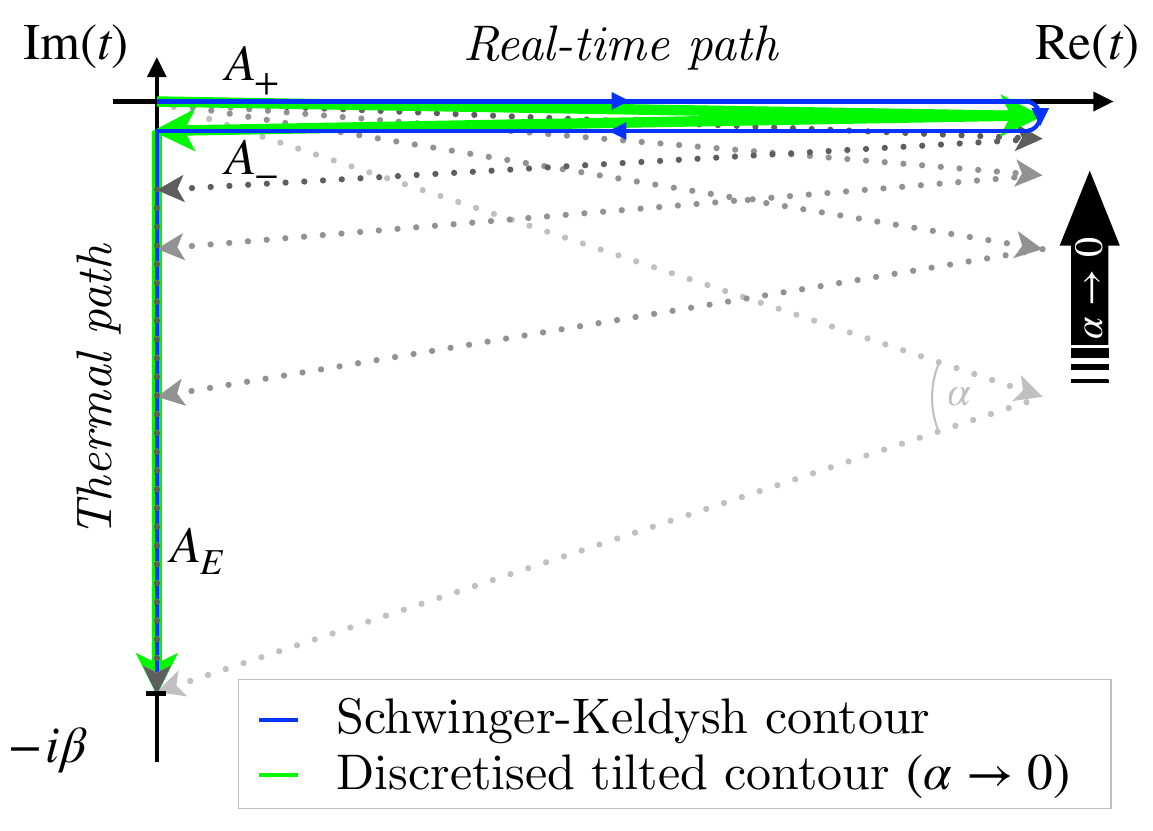}}
\end{figure}

Despite ongoing advances in theoretical development, first principle computations of the evolution of the quark-gluon plasma are still unattainable. This is largely due to the presence of the numerical sign problem when simulating gauge theories in real time. Of particular interest are unequal-time correlation functions of the energy-momentum tensor. These correlations are essential in the determination of the transport coefficients of the quark-gluon plasma such as shear and bulk viscosities or the speed of sound. We can extract these quantities from unequal time correlation functions, which can be expressed in terms of a path integral along the Schwinger-Keldysh (SK) contour visualized in blue in Fig.~\ref{fig:SK}:
\begin{align}
    \label{eq:path_OO}
    \left\langle \mathscr{O}(t) \mathscr{O}(t') \right\rangle
   =\frac{1}{Z}\int \mathcal{D}A_E \, e^{-S_E[A_E]} \int \mathcal{D}A_+\, \mathcal{D}A_-\, e^{i S[A_+,A_-]} \, \mathscr{O}(t) \mathscr{O}(t'),
\end{align}
where $Z$ denotes the partition function and $A_\pm$ are the gauge fields along the forward and backward branches of the SK contour while Euclidean fields $A_E$ are defined on the thermal path whose extent corresponds to the inverse temperature $\beta = 1/T$. We have periodicity $A(t=0)=A(t=-i\beta)$ for thermal equilibrium. In the path integral \eqref{eq:path_OO}, the sign problem becomes evident when we consider that the action is real along the real-time part of the SK contour. This leads to a complex phase $\exp\left(i S[A_+,A_-]\right)$ as the weight factor for which traditional Monte Carlo techniques are inapplicable. In the recent past, new methods and computational strategies have nevertheless been proposed to obtain results \cite{Alexandru:2020wrj}.  

In this conference contribution, we use the complex Langevin (CL) method that is based on stochastic quantization \cite{Parisi:1983mgm}. We present results from our recent studies \cite{Boguslavski:2022dee, Boguslavski:2023unu} for non-Abelian SU($N_c$) gauge theories on complex time contours. As detailed there, we introduce an anisotropic kernel that originates from a careful rederivation of the CL equation and stabilizes the CL dynamics. This enables the first direct lattice calculation of an unequal-time correlation function. In particular, we compute correlations of the magnetic contribution to the energy density in 3+1 dimensions for SU(2) gauge theory in thermal equilibrium.

\section{Theoretical Background}

In this section, we briefly discuss the most important concepts that underlie the complex Langevin method. We further list the most pressing difficulties associated with this technique and how we address them in the simulations of real-time non-Abelian gauge theories. This lays the fundament for our computation of unequal time correlation functions. 

\subsection{Concepts underlying the CL method}
The complex Langevin method is formulated by continuing the physical theory from the real to the complexified manifold as its configuration space. In the case of SU($N_c$) gauge theory, this generalizes the gauge potential to a non-unitary field: 
\begin{align}
    \mathfrak{su}(N_c) \ni A_{\mu}(x) \longrightarrow A_{\mu}(x) \in \mathfrak{sl}(N_c, \mathbb{C}) = \mathfrak{su}(N_c) + i \, \mathfrak{su}(N_c).
\end{align}
We note that this requires the action and all observables to be infinitely differentiable and their power series to converge at every point. 
This poses a strong assumption on the theory and is not satisfied in general. In the presence of fermions, the action is non-holomorphic at the roots of the fermion determinant. Nonetheless, progress has been achieved in QCD at finite density and it has been argued that the aforementioned assumption can be weakened in certain scenarios \cite{Seiler:2017wvd}.

The centerpiece of this method is the CL equation
\begin{align} \label{eq:cle}
    \frac{\partial A_\mu}{\partial \theta}(x, \theta) = i \left. \frac{\delta S_\mathrm{YM}[A(x')]}{\delta A^a_\mu(x)} \right\vert_{A(x) = A(x, \theta)} + \eta^a_\mu(x, \theta),
\end{align}
where $S_\mathrm{YM}$ is the Yang-Mills action and we have added a fifth coordinate to the gauge potential, the Langevin time $\theta$. The Gaussian noise, which makes this equation a stochastic differential equation, is characterized by
\begin{align}
    \langle \eta_\mu^a(x,\theta) \rangle = 0, \quad
    \langle \eta_\mu^a(x, \theta) \eta_\nu^b(x',\theta') \rangle = 2 \delta^{ab} \delta_{\mu\nu} \delta(x-x') \delta(\theta-\theta').
\end{align}

The complex Langevin method is motivated by the correspondence between Fokker-Plank and Langevin equations in the \emph{real}-valued case. This correspondence allows the determination of the probability density, that describes the stochastic process in the limit of $\theta \to \infty$ and is given by its stationary solution \cite{Parisi:1983mgm}. It has been shown that under certain assumptions this connection can be generalized to \emph{complex} stochastic differential equations \cite{Aarts:2009uq}. In accordance, the fields after a `thermalization' Langevin time $\theta_0$ 
are characterized by $\rho[A] \propto \exp\left(i S_{\mathrm{YM}}[A]\right)$. This leads to
\begin{align}
    \langle \mathcal{O} [A] \rangle = \int \mathcal{D}A \, \rho[A] \mathcal{O} [A] \approx \lim\limits_{\theta_0\rightarrow\infty} \frac{1}{\Delta \theta}\int_{\theta_0}^{\theta_0+ \Delta \theta}  d \theta\, \mathcal{O}[A(\theta)],
\end{align}
for the expectation value of an observable $\mathcal{O}$ and sufficiently large sampling interval $\Delta \theta$. 
Note that the difficulties associated with traditional Monte Carlo techniques due to the complex weights are circumvented by sampling gauge configurations from the \emph{complex} stochastic process.

\subsection{Complex Langevin on the lattice}

We numerically solve the complex Langevin equation by discretizing the Yang-Mills action by the Wilson plaquette action,
\begin{align}
    S_\mathrm{W}[U] = \frac{1}{g^2} \sum_{x, \mu \neq \nu} \rho_{\mu\nu} \mathrm{Tr}\left[U_{\mu\nu}(x) - 1\right],
\end{align}
with the lattice spacing dependent anisotropy factors $\rho_{0i}(x) = -a_s / a_{t}(x)$, $\rho_{ij}(x) = \bar a_{t}(x) / a_s$. The gauge link fields and plaquette variables are given by
\begin{align}
    U_{\mu}(x) &\simeq \exp \big[ i g a_\mu(x) A_\mu\big(x+\hat \mu/2\big)\big] \in \mathrm{SL}(N_c, \mathbb{C}) \flowsfromb \mathrm{SU}(N_c), \\ 
    U_{\mu\nu}(x) &= U_{\mu}(x) U_{\nu}(x+\hat\mu) U_{\mu}^{-1}(x+\hat\nu) U_{\nu}^{-1}(x).
\end{align}
The above formulation is analytic and agrees with the SU($N_c$) version for unitary fields.

Note that the temporal lattice spacing $a_{t}(x)$ exhibits time dependence due to the discretization of the complex time contour. For brevity, we write this as a general $x$-dependence to avoid the need for differentiating between temporal and spatial dependencies. We further introduced $\bar a_{t}(x) = (a_{t}(x)+a_{t}(x+\hat{0})/2$ to retain the time-reversal symmetry of the Wilson action also on general complex time contours.

The solution of the CL equation \eqref{eq:cle} is obtained by the Euler-Maruyama scheme in terms of the gauge link formalism:
\begin{align}
    U_{\mu}( x , \theta +  \epsilon) = \exp \left[ i t^a \left( \epsilon\,  \left. \frac{\delta S_\mathrm{W}}{\delta A^a_\mu(x)} \right\vert_{A(x) = A(x, \theta)}  + \sqrt{\epsilon} \eta^a_{\mu}(x, \theta) \right) \right] U_{\mu}(x, \theta).
\end{align}
Details on the derivation of the CL formalism and the drift term on general complex time contours can be found in our recent work \cite{Boguslavski:2022dee}.

\subsection{Instabilities}

Although there have been remarkable advances in the development of the CL method, situations exist where CL is limited in its applicability. The main reason is the emergence of two types of issues: numerical runaways and wrong convergences. The former have been traced back to the instability of numerical schemes \cite{Flower:1986hv, Aarts:2009dg, Alvestad:2021hsi}, which are applied to solve the complex Langevin equation, while the latter are due to the appearance of non-vanishing boundary terms during the evolution and the spectrum of the Fokker-Plank evolution operator \cite{Aarts:2009uq, Scherzer:2018hid, Scherzer:2019lrh, Seiler:2023kes}.
Earlier studies of real-time Yang-Mills theory \cite{Berges:2006xc} have shown that CL simulations suffer from both issues.
Nevertheless, in the past two decades, the CL method has been improved by applying stabilization techniques with the aim of removing these complications.

In this work, we take advantage of the kernel freedom of Langevin equations. This allows the transformation of the drift and the noise term without altering the stationary solution of the stochastic differential equation, but it can drastically improve the stability of the dynamics \cite{Okamoto:1988ru, Okano:1991tz}. In particular, we apply the \emph{anisotropic kernel} that is motivated by a contour parametrization-dependent CL formulation and was derived in our earlier paper \cite{Boguslavski:2022dee}. This kernel effectively rescales the time step of the Langevin evolution for different directions
\begin{align} \label{eq:gamma}
    \begin{split}
        \epsilon \quad \to \quad \epsilon_\mu = \Gamma_\mu \epsilon, \\
        \Gamma_0 = \frac{\vert a_t \vert^2}{a_s^2}, \quad    \Gamma_s = 1.
    \end{split}
\end{align}
In the aforementioned work, we have shown that this modification systematically improves the stability of the CL algorithm by tuning the anisotropy.

Moreover, we utilize the gauge cooling method that uses the gauge invariance of Yang-Mills theory to minimize the non-unitarity of the gauge configurations. This is commonly quantified by the unitarity norm
\begin{align}\label{eq:un}
    F[U] = \frac{1}{4 N_t N_s^3} \sum_{x,\mu} \mathrm{Tr}\left[ (U_{\mu}(x) U_{\mu}^\dagger(x) - \mathbbm 1)^2 \right].
\end{align}
The norm $F[U]$ vanishes for unitary configurations and a large $F[U]$ is often understood as a measure of the instability of complex Langevin simulations. Other than gauge invariance, the effect of gauge cooling profits from the compactness of the original unitary configuration space. Hence, gauge cooling can further be interpreted as a ``quasi-compactification'' of the sampling space. We remark that although it is non-holomorphic, it was shown that gauge-invariant observables are unaffected by this cooling procedure \cite{Nagata:2015uga}.

\subsection{Simulation strategy}

It has been found that the discretized path integral along the Minkowski action leads to an ill-defined temporal continuum limit when the Wilson action is employed \cite{Matsumoto:2022ccq}. It has been therefore suggested to rotate the gauge coupling by a complex phase factor to regulate the integral. In the context of this work, we interpret this phase as a tilt of the real-time contour to effectively regularize the path integral on the lattice. In particular, we tilt the forward and backward time branches towards the Euclidean time direction. This is visualized in Fig.~\ref{fig:SK} by the dotted gray lines, where we hold the real-time extent $t_{\mathrm{max}}$ and the inverse temperature $\beta$ of the contour fixed while decreasing the angle $\alpha$. In addition, this adds a real part to the action of the path integral, thus weakening the severity of the sign problem and allowing for more stable CL simulations. However, we need to extrapolate towards the Schwinger-Keldysh contour (illustrated by a black arrow in Fig.~\ref{fig:SK}), which requires simulations at various tilt angles.

The anisotropic kernel in Eq.~\eqref{eq:gamma} allows us to counteract the instabilities for shrinking tilt angles. This is achieved by increasing the bare anisotropy $a_s/|a_t|$ such that the kernel in conjunction with the gauge cooling technique stabilizes the CL dynamics for the smallest necessary angle to perform an extrapolation $\alpha\to 0$. This anisotropy is imposed for all simulations (at different tilt angles) to obtain sufficient data for the extrapolation. 

\subsection{Correlations of the magnetic energy density}

Transport coefficients of the QGP such as the shear and bulk viscosities or the speed of sound can be expressed in terms of the unequal-time correlation function of the energy-momentum tensor $T_{\mu\nu}$. Closely related to this quantity, we consider the (chromo-)magnetic contribution to the energy density $\frac{1}{2} B^2$. We approximate this quantity by expanding the relation of the field strength tensor and plaquette variables
\begin{align}
    U_{ij}(x) = \exp\left[i a_s^2 F_{ij}(x) + \mathcal{O}(a_s^3) \right].
\end{align}
In terms of cloverleaves, which are averages of four neighboring planar plaquettes,
\begin{align}
\begin{split}
    C_{ij}(x) = \frac{1}{4} [&U_{ij}(x) + U_{i(-j)}(x) + U_{(-i)(-j)}(x) + U_{(-i)j}(x)],
\end{split}
\end{align}
we obtain the magnetic energy density as
\begin{align}
    B^2(x) = \mathrm{Tr}[ F_{ij}(x) F^{ij}(x)] 
    \approx - \frac{1}{a^4_s}\sum_{i, j} \mathrm{Tr}\left\{ \mathcal{P}_A \left[ C_{ij}(x) \right]^2\right\},
\end{align}
where $\mathcal{P}_A$ projects on the anti-hermitian trace-zero part of a given matrix. This expression is used to reduce observable-based cutoff effects compared to the analogous definition using plaquette variables directly. 

In our study, we calculate unequal-time correlations of $\frac{1}{2} B^2$ averaged over the spatial lattice,
\begin{align} \label{eq:C}
    C(t, t') = \frac{1}{4 N_s^3} \sum_\mathbf{x}
    \left\langle B^2(t,\mathbf{x}) B^2(t',\mathbf{x}) \right\rangle.
\end{align}
The choice of this observable is mainly motivated by its improved statistical properties due to the average over $\mathbf x$ and serves as a first study of correlations in Yang-Mills theory in 3+1 dimensions. As the spatial sum is a linear operation, it still allows us to explicitly check different analytic features of the correlation function in thermal equilibrium. In turn, we give evidence for the correctness of our CL simulations footed on this \emph{non-local} observable.

\section{Results}

In this section, we first summarize the numerical setup and model parameters employed in our CL simulations of real-time Yang-Mills theory. These simulations are performed to compute the real-time correlation function of the magnetic energy density $C(t,t')$ given in Eq.~\eqref{eq:C}. We then justify our extrapolation procedure and introduce the nomenclature for different parts of $C(t,t')$. Finally, we show the obtained real-time correlation function and argue that it satisfies a non-trivial consistency relation expected from the continuum theory. For more details, we refer to Ref.~\cite{Boguslavski:2023unu}.

\subsection{Numerical setup}

We simulate SU(2) pure gauge theory on an $N_t\times N_s^3$ lattice with $N_s = 16$ that respects the tilted real-time contour by introducing a time-dependent complex temporal lattice spacing $a_t(x)$. All dimensionful quantities presented in this work are given in units of $a_s$. We impose a lattice anisotropy of $a_s / \vert a_{t} \vert \approx 16$, which suffices to stabilize contours with inverse temperature $\beta /a_s =1/(T a_s)=1$ and real-time extent of $t_\mathrm{max} = 1.5 \beta$ by using the anisotropic kernel \eqref{eq:gamma}. We further assume $g=0.5$ for the gauge coupling in all simulations. In addition, we perform one gauge cooling step with a step size of $0.05$ after each CL update (see Appendix in \cite{Boguslavski:2022dee} for details). We consider tilted time contours with angles $\alpha$ ranging from $\tan(\alpha)=1/3$ to $\tan(\alpha)=1/96$.

We evolve the gauge fields starting at $\theta=0$ with configurations of identity matrices and use an improved Langevin step \cite{Fukugita:1986tg} with a step size of $\epsilon=10^{-4}$. For the numerical evaluation of unequal time correlations, we simulate until the thermalization of the Langevin evolution has been ensured and draw $100$ independent equidistant configurations in the Langevin time interval $\theta=10$ to $20$. To counteract increasingly noisy data for shrinking tilt angles, we also average over $100$ (for $\tan(\alpha)=1/3$) up to $1000$ (for $\tan(\alpha)=1/96$) independent Langevin trajectories. 

\subsection{Extrapolation towards the Schwinger-Keldysh contour}

\begin{figure}[t!]
    \centering
    \begin{subfigure}{.49\textwidth}
      \centering
      \includegraphics[width=0.95\textwidth]{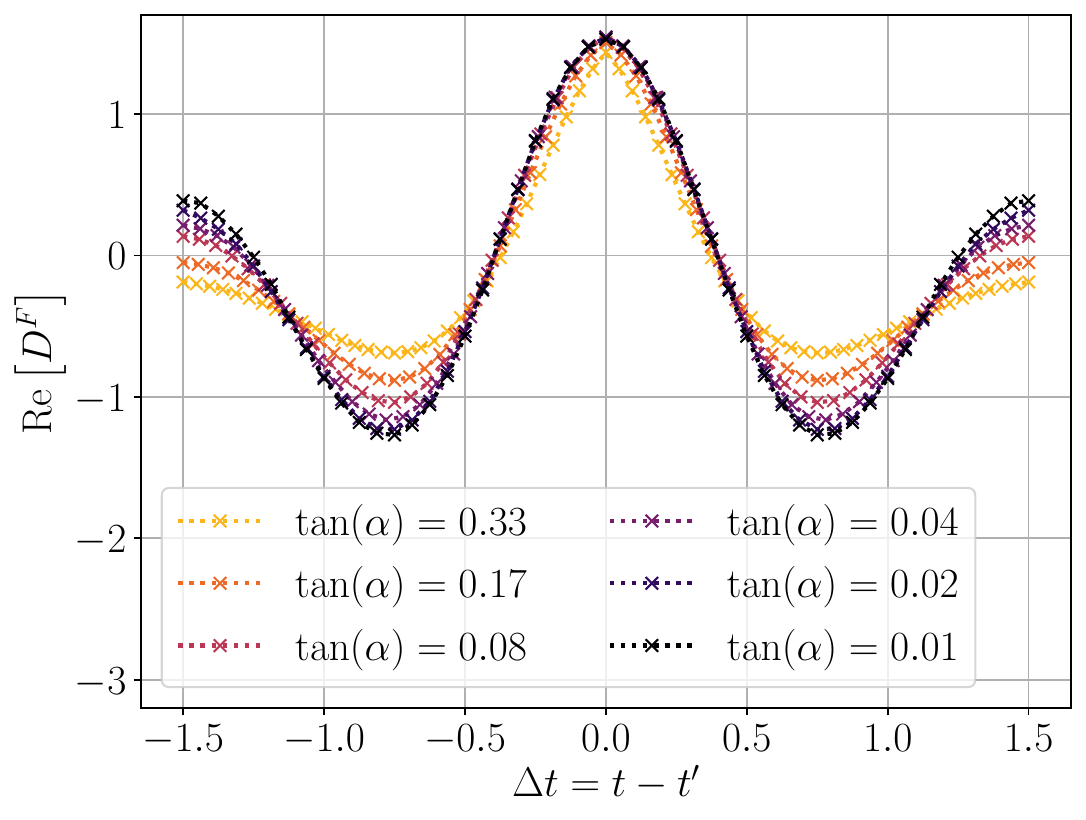}
    \end{subfigure}
    \hfill
    \begin{subfigure}{.49\textwidth}
      \centering
      \includegraphics[width=0.95\textwidth]{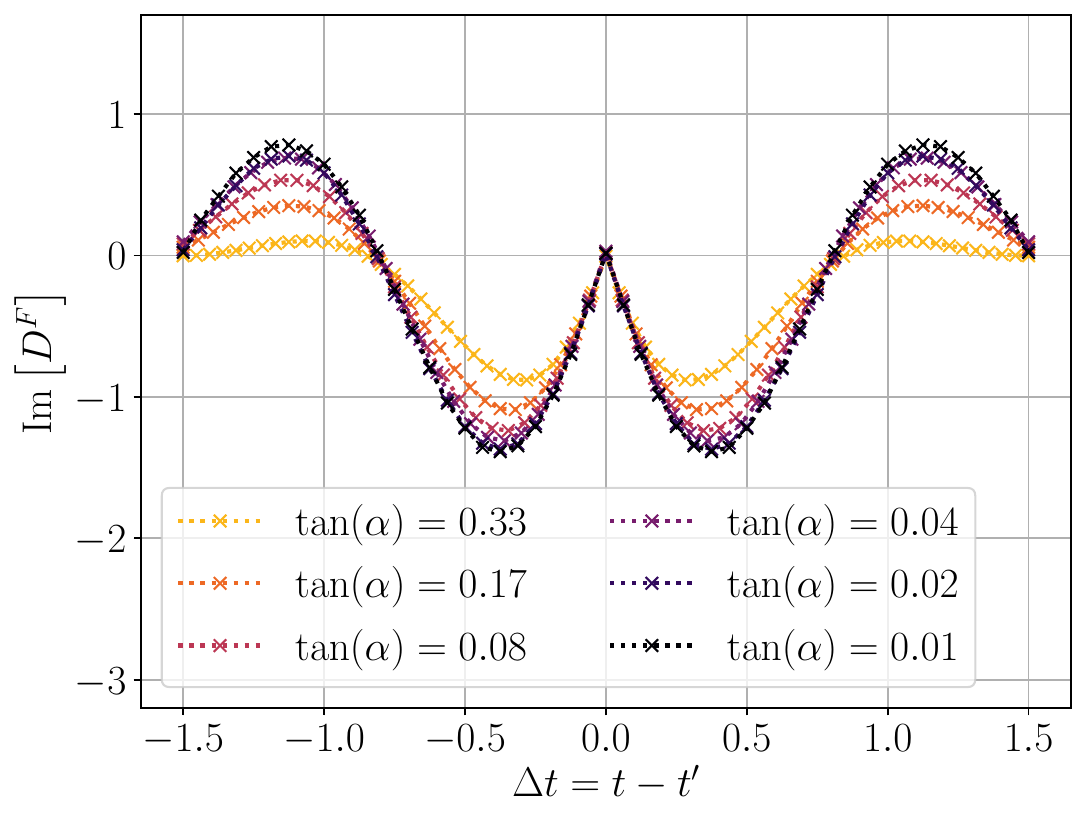}
    \end{subfigure}
    \caption{
    The figure shows the real (\emph{left}) and imaginary (\emph{right}) parts of the Feynman propagator of the magnetic contribution to the energy density. We present this correlation for various tilt angles $\alpha$ for the complex time contour from $\tan(\alpha)=0.33$ to $0.01$. The change of the curve slows down for small $\alpha$ indicating convergence towards the real-time correlation function.
    }
    \label{fig:DF}
\end{figure}

In Fig.~\ref{fig:DF} we show the real (\emph{left panel}) and imaginary (\emph{right panel}) part of the Feynman propagator $D^F$ that is defined as the correlator on the forward time branch
\begin{align}
    D^F(\Delta t = t-t') \equiv
    D^F(t, t') = C(t_+, t'_+).
\end{align}
We show all correlations as functions of the time difference $\Delta t = t-t'$ as the time-translation invariance in thermal equilibrium imposes the independence of the central time coordinate $(t+t')/2$. We observe that the observables converge for shrinking angles.
Notably, we have checked and observed the same behavior independently for the correlations on the backward path as well as the correlations between forward and backward branches. This strongly supports our approach to obtain real-time correlations by extrapolating $\alpha \to 0$.

\begin{figure}[t!]
    \centering
    \begin{subfigure}{.49\textwidth}
      \centering
      \includegraphics[width=0.8\textwidth]{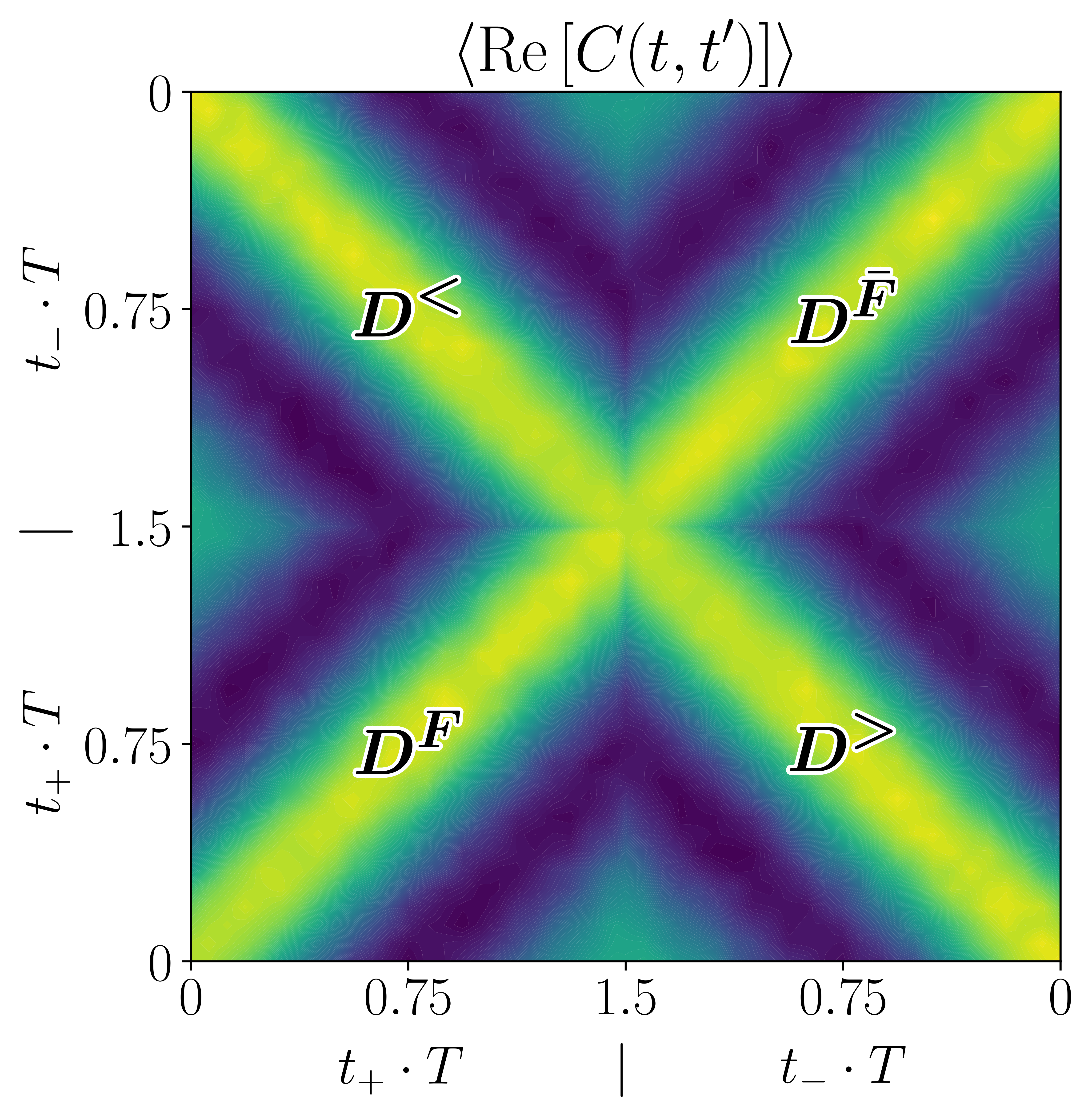}
    \end{subfigure}
    \hfill
    \begin{subfigure}{.49\textwidth}
      \centering
      \includegraphics[width=0.8\textwidth]{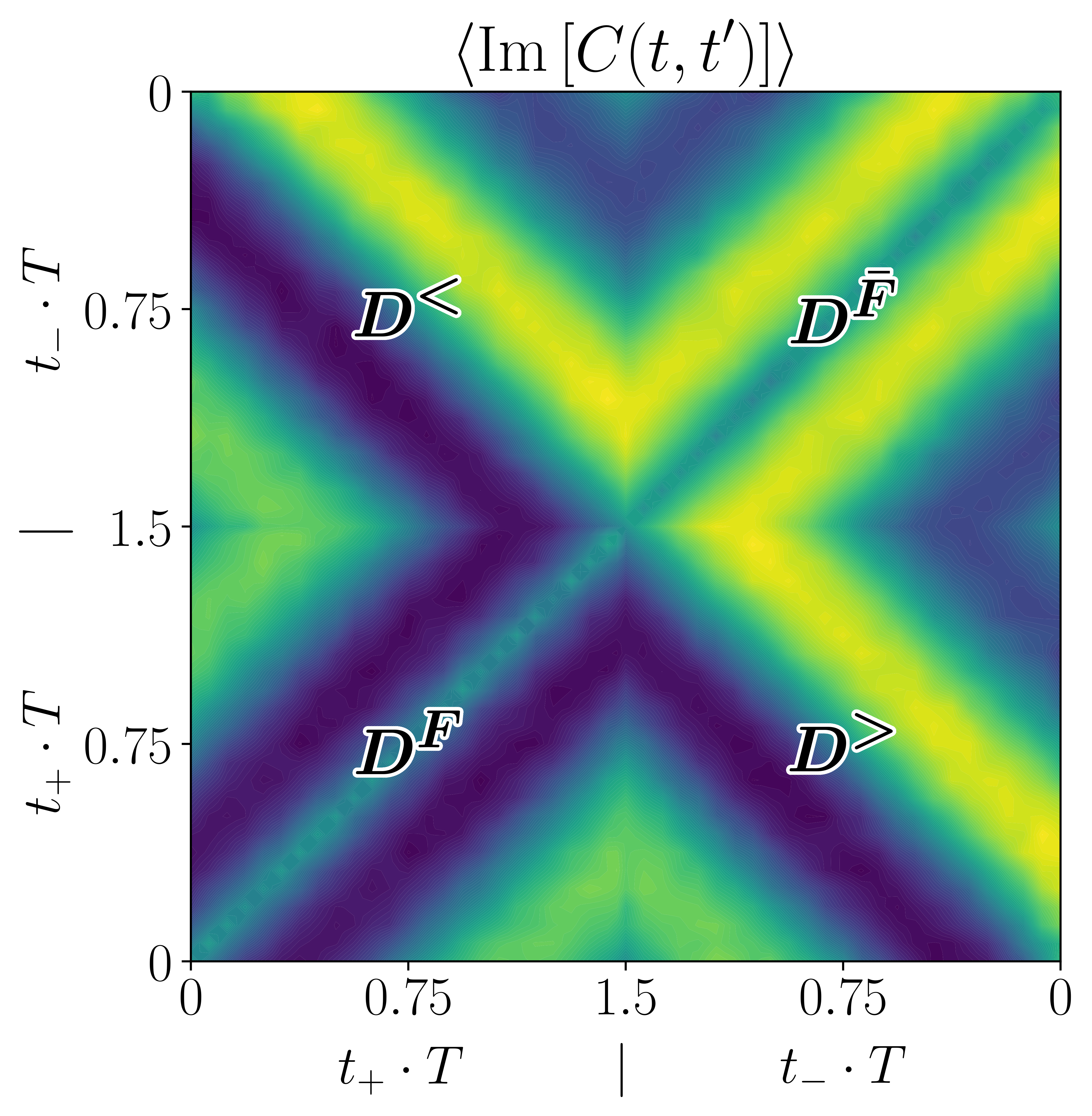}
    \end{subfigure}
    \caption{
    The figure shows the real (\emph{left}) and imaginary (\emph{right}) parts of the correlation function $C(t, t')$ extrapolated towards the SK contour ($\alpha\to0$) restricted to the real-time branches. The axis labels $t_\pm$ indicate the forward and backward path while the labels in the quadrants specify the corresponding correlation function defined in Eq.\ \eqref{eq:correlations}.
    }
    \label{fig:C}
\end{figure}

We use a cubic polynomial extrapolation with respect to $\alpha$ to obtain the correlation of the magnetic energy density on the real-time path. The result is visualized in Fig.~\ref{fig:C}, where the forward and backward paths are indicated by $t_\pm$, respectively. The figure can be divided into four separate quadrants, each corresponding to a propagator
\begin{equation}\label{eq:correlations}
    \begin{aligned}
        D^<(t,t') = C(t_+, t_-'), & &D^{\bar F}(t,t') = C(t_-, t_-'),\\
        D^F(t,t') = C(t_+, t_+'), & &D^>(t,t') = C(t_-, t_+'),
    \end{aligned}
\end{equation}
where $D^F$, $D^{\bar F}$ denote time ordered and anti-time ordered Feynman propagators and $D^>$, $D^<$ are known as the forward and backward Wightman functions. Moreover, the symmetric structure in these Figures stems from the time-translation invariance of the observable. Statistical errors were examined using a bias-corrected ``delete-one'' Jackknife method, confirming their small magnitude. This underscores that the extrapolation is well-behaved.

\subsection{Correspondence between Feynman propagator and Wightman functions}

\begin{figure}
    \centering
    \begin{subfigure}{.49\textwidth}
      \centering
      \includegraphics[width=0.9\textwidth]{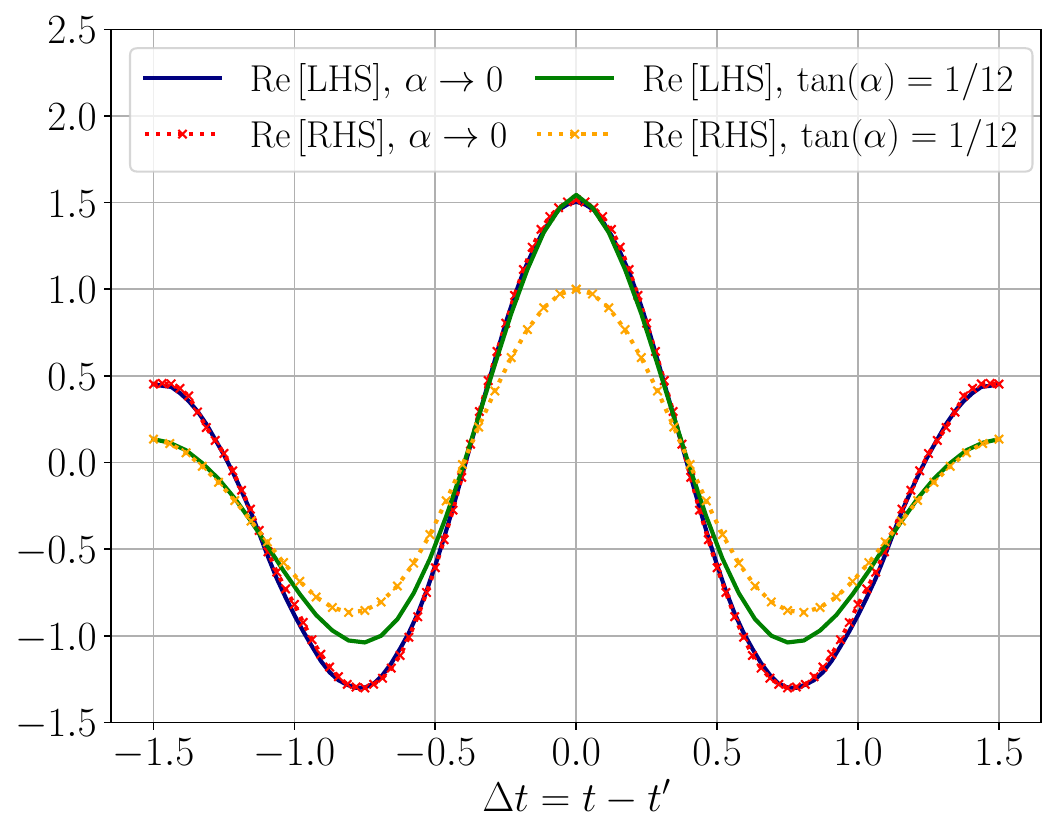}
    \end{subfigure}
    \begin{subfigure}{.49\textwidth}
      \centering
      \includegraphics[width=0.9\textwidth]{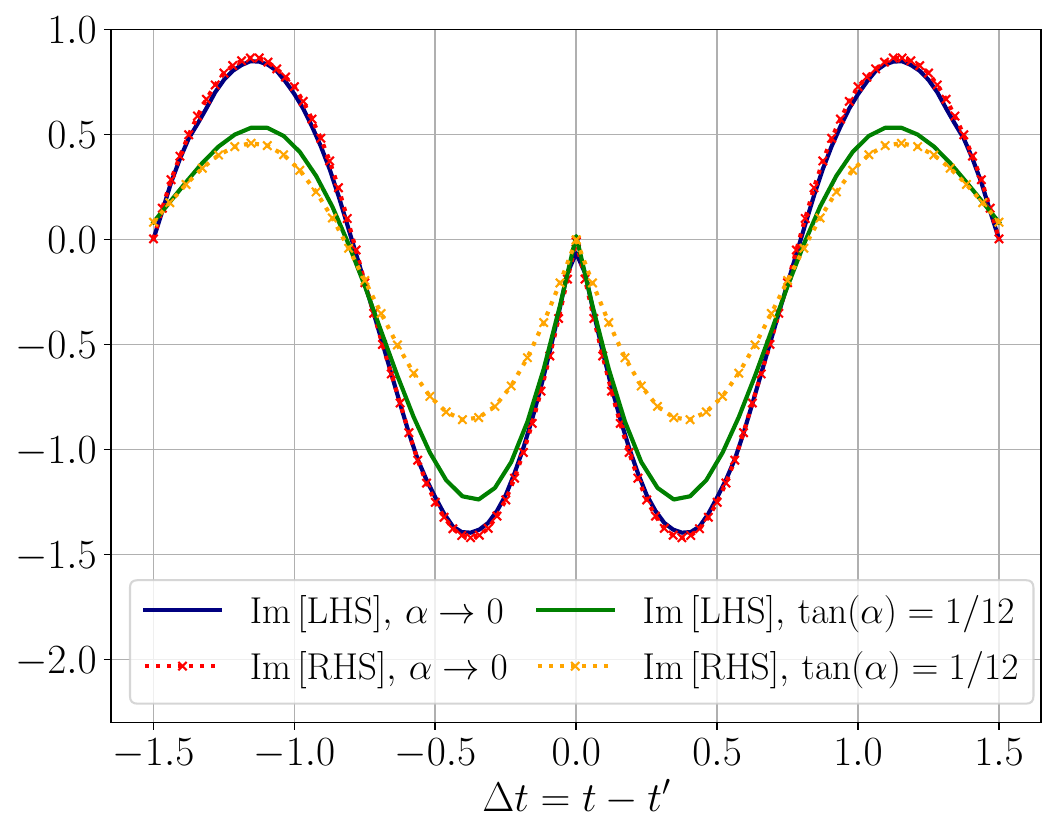}
    \end{subfigure}
    \caption{
        The figure shows the real (\emph{left}) and imaginary (\emph{right}) part of Eq. \eqref{eq:relation}, which relates the Feynman propagator and the Wightman sectors of $C(t,t')$. In both panels, we show that this relation is well-satisfied for the extrapolated data $\alpha\to0$. This relation is not satisfied for finite tilt angles, exemplified by $\tan(\alpha)=1/12$, where both sides of the relation deviate strongly from each other. \label{fig:relation}
    }
\label{fig:test}
\end{figure}

To further confirm our numerical results, we study analytic features that are manifest in quantum field theory. We explicitly check the relation between the Feynman and Wightman propagators, 
\begin{align} \label{eq:relation}
    D^F(t,t') = \Theta(t - t') D^>(t,t') + \Theta(t' - t) D^<(t,t'),
\end{align}
where $\Theta$ is the Heaviside step function.
Our numerical results are presented in Fig.~\ref{fig:relation} where we show the left- and right-hand side of Eq.~\eqref{eq:relation} by separate lines. We show the correlations based on the extrapolated data $\alpha\to 0$ and at finite tilt angle $\tan(\alpha)=1/12$. Strikingly, the relation is only satisfied to very good accuracy for the former. At a finite tilt angle, we see large deviations between both sides of the relation. This further supports the correctness of our extrapolation strategy.

\section{Conclusion}

In this proceeding, we have presented the first direct computation of real-time correlation functions in 3+1 dimensional SU(2) Yang-Mills theory, summarizing our results from \cite{Boguslavski:2023unu}. These were achieved by utilizing the CL method based on our recent work \cite{Boguslavski:2022dee}. Despite the presence of a hard sign problem in the calculation of real-time observables, our strategy of combining the anisotropic kernel with the gauge cooling method allows us to perform stable CL simulations, which enables the extraction of real-time observables.

In particular, we have calculated the correlation function of the magnetic contribution to the energy density on discretized Schwinger-Keldysh contours with a real-time extent of $t_{\mathrm{max}} = 1.5\beta$ and inverse temperature $\beta/a_s=1$. The investigation of non-trivial consistency relations shows the correctness of our results and supports the applicability of CL to real-time simulations.

In future work, one of our goals is to increase the real-time extent $t_{\mathrm{max}}$ and coupling $g$ while keeping our simulations stable. This is needed to gain a detailed understanding of the correlation function in frequency space. This may culminate in the calculation of transport coefficients and spectral functions in the future, which are of high phenomenological interest to heavy-ion collision and quark-gluon plasma research.

We note that a renormalization procedure is currently missing for real-time lattice simulations and was not conducted in this work. This is needed to extract physical quantities for such simulations. Our advances in the field and other recent progress in the realm of scalar field theory \cite{Alvestad:2023jgl} underscore the need for a systematic approach to the determination of a scale in real-time lattice simulations.

\begin{acknowledgments}
    This research was funded by the Austrian Science Fund (FWF) under the projects P~34455, P~34764 and \mbox{W~1252}. The computational results presented have been achieved using the Vienna Scientific Cluster (VSC).
\end{acknowledgments}

\bibliographystyle{JHEP}
\bibliography{ref}

\providecommand{\href}[2]{#2}\begingroup\raggedright\begin{thebibliography}{10}

\bibitem{Alexandru:2020wrj}
A.~Alexandru, G.~Basar, P.F.~Bedaque and N.C.~Warrington, \emph{{Complex paths around the sign problem}}, \href{https://doi.org/10.1103/RevModPhys.94.015006}{\emph{Rev. Mod. Phys.} {\bfseries 94} (2022) 015006} [\href{https://arxiv.org/abs/2007.05436}{{\ttfamily 2007.05436}}].

\bibitem{Parisi:1983mgm}
G.~Parisi, \emph{{ON COMPLEX PROBABILITIES}}, \href{https://doi.org/10.1016/0370-2693(83)90525-7}{\emph{Phys. Lett. B} {\bfseries 131} (1983) 393}.

\bibitem{Boguslavski:2022dee}
K.~Boguslavski, P.~Hotzy and D.I.~M\"uller, \emph{{Stabilizing complex Langevin for real-time gauge theories with an anisotropic kernel}}, \href{https://doi.org/10.1007/JHEP06(2023)011}{\emph{JHEP} {\bfseries 06} (2023) 011} [\href{https://arxiv.org/abs/2212.08602}{{\ttfamily 2212.08602}}].

\bibitem{Boguslavski:2023unu}
K.~Boguslavski, P.~Hotzy and D.I.~M\"uller, \emph{{Real-time correlators in 3+1D thermal lattice gauge theory}},  \href{https://arxiv.org/abs/2312.03063}{{\ttfamily 2312.03063}}.

\bibitem{Seiler:2017wvd}
E.~Seiler, \emph{{Status of Complex Langevin}}, \href{https://doi.org/10.1051/epjconf/201817501019}{\emph{EPJ Web Conf.} {\bfseries 175} (2018) 01019} [\href{https://arxiv.org/abs/1708.08254}{{\ttfamily 1708.08254}}].

\bibitem{Aarts:2009uq}
G.~Aarts, E.~Seiler and I.-O.~Stamatescu, \emph{The {Complex Langevin} method: When can it be trusted?}, \href{https://doi.org/10.1103/PhysRevD.81.054508}{\emph{Phys. Rev. D} {\bfseries 81} (2010) 054508} [\href{https://arxiv.org/abs/0912.3360}{{\ttfamily 0912.3360}}].

\bibitem{Flower:1986hv}
J.~Flower, S.W.~Otto and S.~Callahan, \emph{{Complex Langevin Equations and Lattice Gauge Theory}}, \href{https://doi.org/10.1103/PhysRevD.34.598}{\emph{Phys. Rev. D} {\bfseries 34} (1986) 598}.

\bibitem{Aarts:2009dg}
G.~Aarts, F.A.~James, E.~Seiler and I.-O.~Stamatescu, \emph{{Adaptive stepsize and instabilities in complex Langevin dynamics}}, \href{https://doi.org/10.1016/j.physletb.2010.03.012}{\emph{Phys. Lett. B} {\bfseries 687} (2010) 154} [\href{https://arxiv.org/abs/0912.0617}{{\ttfamily 0912.0617}}].

\bibitem{Alvestad:2021hsi}
D.~Alvestad, R.~Larsen and A.~Rothkopf, \emph{{Stable solvers for real-time Complex Langevin}}, \href{https://doi.org/10.1007/JHEP08(2021)138}{\emph{JHEP} {\bfseries 08} (2021) 138} [\href{https://arxiv.org/abs/2105.02735}{{\ttfamily 2105.02735}}].

\bibitem{Scherzer:2018hid}
M.~Scherzer, E.~Seiler, D.~Sexty and I.-O.~Stamatescu, \emph{Complex {Langevin} and boundary terms}, \href{https://doi.org/10.1103/PhysRevD.99.014512}{\emph{Phys. Rev. D} {\bfseries 99} (2019) 014512} [\href{https://arxiv.org/abs/1808.05187}{{\ttfamily 1808.05187}}].

\bibitem{Scherzer:2019lrh}
M.~Scherzer, E.~Seiler, D.~Sexty and I.O.~Stamatescu, \emph{{Controlling Complex Langevin simulations of lattice models by boundary term analysis}}, \href{https://doi.org/10.1103/PhysRevD.101.014501}{\emph{Phys. Rev. D} {\bfseries 101} (2020) 014501} [\href{https://arxiv.org/abs/1910.09427}{{\ttfamily 1910.09427}}].

\bibitem{Seiler:2023kes}
E.~Seiler, D.~Sexty and I.-O.~Stamatescu, \emph{{Complex Langevin: Correctness criteria, boundary terms and spectrum}},  \href{https://arxiv.org/abs/2304.00563}{{\ttfamily 2304.00563}}.

\bibitem{Berges:2006xc}
J.~Berges, S.~Borsanyi, D.~Sexty and I.O.~Stamatescu, \emph{{Lattice simulations of real-time quantum fields}}, \href{https://doi.org/10.1103/PhysRevD.75.045007}{\emph{Phys. Rev. D} {\bfseries 75} (2007) 045007} [\href{https://arxiv.org/abs/hep-lat/0609058}{{\ttfamily hep-lat/0609058}}].

\bibitem{Okamoto:1988ru}
H.~Okamoto, K.~Okano, L.~Schulke and S.~Tanaka, \emph{{The Role of a Kernel in Complex Langevin Systems}}, \href{https://doi.org/10.1016/0550-3213(89)90526-9}{\emph{Nucl. Phys. B} {\bfseries 324} (1989) 684}.

\bibitem{Okano:1991tz}
K.~Okano, L.~Schulke and B.~Zheng, \emph{Kernel controlled complex {Langevin} simulation: Field dependent kernel}, \href{https://doi.org/10.1016/0370-2693(91)91111-8}{\emph{Phys. Lett. B} {\bfseries 258} (1991) 421}.

\bibitem{Nagata:2015uga}
K.~Nagata, J.~Nishimura and S.~Shimasaki, \emph{{Justification of the complex Langevin method with the gauge cooling procedure}}, \href{https://doi.org/10.1093/ptep/ptv173}{\emph{PTEP} {\bfseries 2016} (2016) 013B01} [\href{https://arxiv.org/abs/1508.02377}{{\ttfamily 1508.02377}}].

\bibitem{Matsumoto:2022ccq}
N.~Matsumoto, \emph{{Comment on the subtlety of defining real-time path integral in lattice gauge theories}}, \href{https://doi.org/10.1093/ptep/ptac106}{\emph{PTEP} {\bfseries 2022} (2022) 093B03} [\href{https://arxiv.org/abs/2206.00865}{{\ttfamily 2206.00865}}].

\bibitem{Fukugita:1986tg}
M.~Fukugita, Y.~Oyanagi and A.~Ukawa, \emph{{Langevin Simulation of the Full {QCD} Hadron Mass Spectrum on a Lattice}}, \href{https://doi.org/10.1103/PhysRevD.36.824}{\emph{Phys. Rev. D} {\bfseries 36} (1987) 824}.

\bibitem{Alvestad:2023jgl}
D.~Alvestad, A.~Rothkopf and D.~Sexty, \emph{{Lattice real-time simulations with learned optimal kernels}},  \href{https://arxiv.org/abs/2310.08053}{{\ttfamily 2310.08053}}.

\end{thebibliography}\endgroup

\end{document}